\documentstyle[12pt]{article}

\newcommand{\be}{\begin{equation}}
\newcommand{\ee}{\end{equation}}
\newcommand{\bea}{\begin{eqnarray}}
\newcommand{\eea}{\end{eqnarray}}
\newcommand{\nn}{\nonumber\\}
\begin{document}
\hfill{HYUPT-94/12}
\vskip 2.5cm
\begin{center}

{\large{\bf Characteristics of QCD Phase Transitions\\ in an Extended Skyrme
Model on $S^3$}
\vskip 1.5cm
{\underline{Joon Ha Kim}, Sooman Yee\footnote{Permanent address: Dept. of
Physics,
Soonchunhyang Univ., Onyang P.O.B.97, Chungnam 336-600 Korea} and
Hyun Kyu Lee\\ \vskip .5cm Dept. of Physics, Hanyang University,
Seoul 133-791 Korea\\ \vskip .5cm e-mail adress: kjh@hepth.hanyang.ac.kr}}

\vskip 2cm

\begin{quote}
\item[] \hskip 1.5cm We study the characteristics  of the QCD phase
transitions in dense hadronic matter using the Skyrme model constructed on
$S^3$.
We  find numerically the localized solutions on $S^3$ using the
extended Skyrme model which implements correctly the scale symmetry
of QCD.  The transition from the localized phase to the
delocalized phase  is found to be
of first order  at the critical radius of the hypersphere, $L_c$.
The chiral restoration and the gluon decondensation also take
place  at the same critical size.

\end{quote}
\end{center}

keywords: extended Skyrme model, hypersphere, phase transition,
          scale symmetry, chiral symmetry restoration, gluon decondensation.
\newpage

The Skyrme model\cite{skyrme} describing baryons as topological solitons
(skyrmion) arising from the
effective meson theory of QCD  in the large $N_c$-limit  has been proved
to be quite successful  in  calculating the various physical properties
of baryons which are in fair quantitative agreements with experimental data.
Given the success of the Skyrme model, there have been also interesting
developments to investigate the behavior of hadronic matter at  higher
 density.
One of the interesting methods, pioneered by  Manton and
Ruback\cite{maru,manton} is to put one skyrmion  on a hypersphere $S^3(\rho)$
with
three dimensional volume $2\pi^2 \rho^3$. Although the obvious disadvantage of
this method
is that the physical space is not a hypersphere  with curvature,
this approach has been adopted to investigate the problems of dense matter in
the Skyrme model motivated  by the observation that  it gives, in a simple
frame work,  similar results as
  in the lattice skyrmion approach\cite{kleba} and also  by its mathematical
simplicity. The hypersphere approach has been studied not only in the Skyrme
model\cite{jwc,hypsky} but also recently in the Nambu-Jona-Lasino
model\cite{forkel}.

The skyrmion on the hypersphere of infinitely large
radius is well localized so that it has the same properties as an isolated
skyrmion in the flat space.
 The finite size solution which is well localized is possible due to the
presence of the interaction term(Skyrme term or higher derivative terms),
which prevents it from being collapsed.
However,
by shrinking the radius of the hypersphere, the effect of
interaction term becomes much stronger than the kinetic term  because of the
curvature of the hypersphere.
Therefore, it is possible that for the  radius smaller than a certain critical
 radius $\rho_c$, the localization
mechanism due to the kinetic term becomes inactive and the skyrmion will
spread out over the whole
hypersphere uniformly. Manton \cite{manton} interpreted it as a
signature for  the quark-hadron phase
transition. Moreover, in relation to the chiral symmetry, the
observation\cite{jwc} of vanishing order parameter
$\langle \sigma \rangle_0$ at the critical density leads to the
interpretation that it would be a signature for chiral symmetry  restoration.

In QCD, one of the
important order parameter, is the gluon condensate
 $\langle 0| F^{\mu\nu}F_{\mu\nu}|0\rangle$, which breaks the scale invariance
in the hadron phase of QCD.  The simplest way to implement this scale symmetry
 of QCD
is to introduce a scalar field, $\chi$\cite{ella}, a dilaton or a glueball
field in the effective Lagrangian.
 The scale anomaly of QCD  in low energy
physics is provided by the effective potential of $\chi$ of the
following generic form,
\be
\frac{B}{4} \chi^4 (\ln \chi^4 -1),\label{xpotential}
\ee
where $B$ is related to the bag constant. Eq.(\ref{xpotential}) has its
minimum at nonzero
$\chi_0=\langle \chi \rangle_0$ which
correctly reproduces the scale anomaly
\cite{schechter}, and it is related to the gluon condensate,
\be
\chi_0^4= -\frac{\beta(g)}{2g} \langle 0| F^{\mu\nu}F_{\mu\nu}|0\rangle
\ee
  Recently
the scale symmetry property of QCD in the effective Lagrangian has been proved
 to be very useful in
discussing  the QCD phase
transition\cite{caelol} at finite temperature and finite density and also in
investigating the medium effect on the parameters in the effective
Lagrangian\cite{rho}.
Gomm, et al\cite{gomm} studied the bag formation in the Skyrme model
using the effective  Lagrangian in flat space, $R^3$, defined as

\be
{\cal L} = \frac{f^2_{\pi}}{4} \chi^2 Tr L_{\mu} L^{\mu}
           + \frac{1}{32 e_s} Tr [L_{\mu} L_{\nu}]^2
	   -\frac{C}{2} \partial_{\mu} \chi \partial^{\mu} \chi
	   -\frac{B}{4} \chi^4 (\ln \chi^4 -1),\label{la}
\ee
which has the correct QCD scaling law.
The glueball field
$\chi$ is normalized as $\chi/\chi_0=1$, $C$ and $B$ are related to $\chi_0^2$
and the bag constant($B_B = B {\chi_0}^4/4$) respectively.
Some years ago, Reinhardt and Dang\cite{reda} investigated the behavior of the
skyrmion in the dense baryonic matter using the same Lagrangian on $S^3$.
They found that there is a phase transition into delocalized  Skyrmion
triggered by the vanishing gluon condensate. But, it is not easy to
read out the order of the phase transition and the
critical radius from their calculations.
In this work,  we carry out  the
numerical analysis in more detail to investigate the characteristics of phase
transition, particularly to see whether it is first or  second order
phase transition.

We use the hedgehog ansatz for the skyrmion field
configuration on $S^3$,
\be
U=\exp[i \vec{\tau} \cdot \hat{r}(\theta,\phi) f(\mu)],\label{hedgehog}
\ee
where $(\mu,\theta,\phi)$ are polar coordinates on $S^3(\rho)$ with $0 \leq
\mu,
\theta \leq \pi$ and $0 \leq \phi \leq 2\pi$ and  the metric is given by
\be
ds^2 = \rho^2 (d\mu^2+\sin^2 \mu d\theta^2 + \sin^2 \mu \sin^2 \theta d\phi^2).
\ee
 The static energy of the system can be obtained from eq.(\ref{la}) using
hedgehog ansatz for chiral field and classical field for $\chi(\mu)$,
\bea
E&=& - \int \sqrt{g} \,dV {\cal L}, \nn
&=& \frac{f_{\pi}}{2 e_s} ({\bar{E}}_2 +{\bar{E}}_4 +{\bar{E}}_{\chi}), \nn
{\bar{E}}_2&=& 4 \pi L \int^{\pi}_{0}\,d\mu \sin^2 \mu {\chi}^2
({f^\prime}^2 + 2 \frac{\sin^2 f}{\sin^2 \mu}), \nn
{\bar{E}}_4&=& \frac{4 \pi}{L} \int^{\pi}_{0}\,d\mu \sin^2 f (2 {f^\prime}^2
+\frac{\sin^2 f}{\sin^2 \mu}), \nn
{\bar{E}}_{\chi}&=& 8 \pi e^2_s L \int^{\pi}_{0}\,d\mu \sin^2 \mu
\{ \frac{\bar{C}}{2} {\chi^\prime}^2
+ L^2 [\frac{\bar{B}}{4} \chi^4 (\ln \chi^4 -1)+\frac{\bar{B}}{4}] \},
\eea
where $
\bar{C}= C/(e_s f_{\pi})^2$, $\bar{B}=B/(e_s f_{\pi})^4
$ and $L$ is the dimensionless radius defined as $L=e_s f_{\pi} \rho$.

Extremizing
the static energy, we get the Euler equations for $\chi(\mu)$ and $f(\mu)$,
\bea
f^{\prime \prime} (L \chi^2 + \frac{2}{L}\frac{\sin^2 f}{\sin^2 \mu})
+2 L f^\prime(\chi \chi^\prime + {\chi}^2 \cot \mu)  \nn
-\frac{\sin 2 f}{\sin^2 \mu} [L {\chi}^2 +
\frac{1}{L} (\frac{\sin^2 f}{\sin^2 \mu} - {f^\prime}^2)]=0,\label{feq}\\
\chi^{\prime \prime} +2 \cot \mu \chi^\prime - \frac{1}{\bar{c}}
[4 L^2 \bar{B} \chi^3 \ln \chi + \frac{1}{e_s^2} \chi ( {f^\prime}^2
+ 2 \frac{\sin^2 f}{\sin^2 \mu})]=0.\label{chieq}.
\eea
where the chiral profile function and the glueball field satisfy
the boundary conditions, for the $B=1$ baryon,
\be
f(0)=0,\,\,\,f(\pi)=\pi,\,\,\,\chi^\prime(0)=0,\,\,\,\chi^\prime(\pi)=0.
\ee

One can easily see that the solutions of the identity map $f(\mu)=\mu$ with
$\chi(\mu)=0$ are the solutions of the above
Euler equations for any radius, $L$. They are the delocalized
solutions with no bag formation\cite{gomm}.

We proceed to find numerical solutions with a typical parameter
set\cite{gomm}:
\be
e_4=4.5,\,\,\,f_{\pi}=93MeV,\,\,\,C=3.29\times
10^2(MeV)^2,\,\,\, B=(220MeV)^4/e.
\ee
We can find two numerical solutions which are localized in
addition to the identity map solution above mentioned. In Fig. 1,
We plot the static energies of the solutions with respect to the radius of
the hypersphere, $L$
It is
found that the two set of the localized solutions(curves (a) and (c) in
Fig. 1)
are possible for  $L > L_t(\cong 1.9)$.
One set of the solutions (curve (c) in Fig. 1) is not of our interest,
Since we are interested in the lower energy configurations.
At $L= L_c(\cong 1.9)$ the energy of the identity map
solutions(curve(b)) is equal to the energy of the localized
solutions(curve (a)).
For a smaller size of the hypersphere, $L < L_c$,
the identity map solutions(delocalized phase) are the lower energy
solutions while for the larger hypersphere, $L > L_c$, the localized
solutions are the minimum energy solutions. Hence there is a phase
transition from the localized to the delocalized
phase at $L_c$, which is of first order.
The characteristic of the phase transition
shown in Fig.1 is quite different from the results without $\chi$ field in the
effective Lagrangian, where the phase
 transition is smooth and found to be of second order($L_c=\sqrt{2}$).
Also the changes of the
configurations  of $f(\mu)$ and $\chi(\mu)$ with
$L$ are  found out to be abrupt at $L_c$ as shown in Fig. 2(a) and 2(b)
respectively. Since the delocalized solutions characterized by $\chi(\mu)=0$
everywhere correspond to vanishing gluon condensate, the phase
transition at $L_c$ from
the localized phase with nonvanishing $\chi(\mu)$ at $\mu = \pi$
to delocalized phase implies a discontinuous change of the gluon
condensate at $L_c$.  These
results can be considered as   evidences of the first order phase
transition\cite{wirzba} triggered by the glueball field.

The order parameter, $\langle \sigma \rangle_0$, which is
defined by
\be
\langle \sigma \rangle_0 = -\frac{2}{\pi} \int^\pi_0 d\mu \sin^2\mu
\cos f(\mu).\label{sigma}
\ee
has been introduced as a useful quantity which effectively
measures the chiral symmetry breaking. As shown in Fig.3, one
can see the discontinuous change of the
order parameter $\langle \sigma \rangle_0$ at $L_c$. This
suggests that the
chiral symmetry is restored at the same radius $L_c$ and the
phase transition is of first order.

Some remarks are in order for the localized
solutions. It is found
that the localized solutions(the curves (a) and (c) in Fig.1)
merge at $L_t$ and the localized solutions are  no more
possible for $L < L_t$ in numerical
calculation. In order to see whether it is an numerical
artifact, we look into the gradient of the profile function
at $\mu=0$, $f^\prime(0)$ as a function of $L$. As shown in Fig. 4,
$f^\prime (0)$'s of the two
localized solutions are smoothly joined at $L_t$ and
continuously reverse their slopes respect to $L$,
$ \frac{d f^\prime (0)}{d L}$.  This observation provides at
least an partial explanation of the merging behavior of the two set of the
localized solutions and the absence of localized
solutions for $L < L_t$  other than the
identity map solutions.

It is also interesting to note that our analyses indicate that
the metastable baryonic state, (a) in Fig. 1,  may possibly exist
between $L_t < L < L_c$, since the energy of this state is not
much bigger than the identity map solution (b) in Fig. 1.

In conclusion, in the frame work of the Skyrme model on $S^3$
with scale symmetry implemented, we find the
evidence for the first order phase transition   at the
critical radius of the hypersphere, $L_c$; from the localized phase to
the delocalized phase, the chiral restoration and the gluon decondensation.

This work is supported in part by KOSEF under Grant No. 94-1400-04-01-3
and in part by Ministry of Education(BSRI-94-2441)

\newpage

\hskip 5cm {\large Figure Captions}
\vskip .5cm
Fig. 1 The static energies with respect to the dimensionless radius,
$L$ for two set of the
localized numerical solutions, (a) and (c), and for the identity map
solutions, (b).
\vskip .5cm
Fig. 2 The numerical solutions of the chiral profile function $f(\mu)$, (a)
and the glueball field $\chi(\mu)$, (b).
\vskip .5cm
Fig. 3 The order parameter $ \langle \sigma\rangle_0$. The dahed line is for
a model without glue ball field and the solid line for
 the extended model which is implemented by  the glueball field.
\vskip .5cm
Fig. 4 The slops of the chiral profile function at $\mu = 0$, $f^\prime(0)$ :
The curve (a) corresponds
to the lower energy localized solutions, the curve (a) in Fig. 1,
and the curve (c) corresponds to the higher energy localized solutions,
the curve (c) in Fig. 1.

\end{document}